\documentclass[12pt]{article}
\usepackage{a4wide}
\usepackage{latexsym}
\usepackage{epsfig}
\usepackage{subfigure}
\usepackage{axodraw}
\usepackage{psfrag}
\usepackage{amsmath}
\newcommand{\be}{\begin{equation}}
\newcommand{\ee}{\end{equation}}
\newcommand{\bea}{\begin{eqnarray}}
\newcommand{\eea}{\end{eqnarray}}

\newcommand{\bv}[1]{\boldsymbol{#1}}

\begin{document}


\begin{flushright}
NIKHEF/2005-031\\
BI-TP 2005/53
\end{flushright}

{\begin{center}
\large
\bf
Electromagnetic vertex function of the pion at $T > 0$
\end{center}
\vspace{0.5cm}
{\begin{center}
J. van der Heide$^*$, E. Laermann$^\dag$\\
Fakult\"at f\"ur Physik, Universit\"at Bielefeld, D-33615 Bielefeld, Germany\\
\vspace{0.5cm}
J.H. Koch$^\ddag$\\
National Institute for Nuclear Physics and High-Energy Physics (NIKHEF),\\
1009 DB Amsterdam, The Netherlands\\
and\\
Institute for Theoretical Physics, University of Amsterdam, Valckenierstraat 65,\\
1018 XE Amsterdam, The Netherlands\\
\end{center}
\vspace{2cm}
\begin{center}
{\bf Abstract}\\
\end{center}
The matrix element of the electromagnetic current between pion states is
calculated in quenched lattice QCD at a temperature of $T = 0.93\:T_c$.  The
nonperturbatively improved Sheikholeslami-Wohlert action is used together with
the corresponding ${\cal O}(a)$ improved vector current. The electromagnetic
vertex function is extracted for pion masses down to $360\;{\rm MeV}$ and
momentum transfers $Q^2 \le 2.7\: {\rm GeV}^2$.
\vspace{7cm}
\\
$^*$Electronic address: jan@physik.uni-bielefeld.de\\
$^\dag$Electronic address: edwin@physik.uni-bielefeld.de\\
$^\ddag$Electronic address: justus@nikhef.nl\\

\newpage
\section{Introduction}

The theory governing the strong force, QCD, predicts distinctive changes in the
behaviour of hadronic matter when exposed to extreme conditions, \textit{i.e.}
at high temperature and/or at high baryon density. It is the aim of large
experimental programs to create matter in such an environment and to study its
properties. In order to interpret the experimental findings, one of the
important questions to be answered is to quantify how properties of hadrons
produced in heavy ion collisions are changed in a hot medium.

In this paper we investigate possible changes in the internal structure of
hadrons at a temperature below the critical temperature $T_c$, \textit{i.e.} we
look for precursors of the phase transition for the pion.
More specifically, we use lattice QCD to
calculate spatial two- and three-point functions for a pion,
to extract the pion 'form factor' at finite temperature.

Finite temperature form factors of the pion have been considered in the
framework of a variety of theoretical approaches.  An increase in the charge
radius with increasing temperature was found in the Nambu-Jona Lasinio model
\cite{Schulze:1994fy}.  Dominguez {\it et al.} \cite{Dominguez:1996kf} used a
finite energy QCD sum rule to show that the charge radius of the pion increases
with $T$ and diverges at some critical temperature.  An extensive study of the
pion electromagnetic form factors at finite $T$ was undertaken by Song and Koch
\cite{Song:1996dg} in the time-like region. Using an effective chiral Lagrangian
to one loop, the form factor at $T > 0$ was found to be reduced in magnitude, in
particular in the vector meson dominance region. Similar results were recently
obtained by Nicola {\it et al.} \cite{GomezNicola:2004gg} who work in chiral
perturbation theory to one loop. They calculated the form factor in the
space-like region and found that the charge radius initially stays constant, but
then increases at higher temperature.

There have already been several lattice QCD investigations of some properties of
a pion embedded in a heat bath.  A recent review of these studies, both quenched
and unquenched, can be found in Ref.~\cite{Karsch:2003jg}.  Most of the work
considered spatial correlators and extracted the so called 'screening masses'
and spatial 'Bethe-Salpeter wave functions' of the pion.  The general conclusion
was that for temperatures below the phase transition temperature, $T_c$, there
are no significant differences between the finite temperature 'screening' mass
and the free pion 'pole' mass, extracted at $T = 0$. Furthermore, the
'Bethe-Salpeter wave functions' extracted from spatial two-point correlators
at $T < T_c$ were very similar to those at $T=0$. However, as subsequent
calculations \cite{vanderHeide:2003ip, vanderHeide:2003kh} of the pion form
factor at $T=0$ have shown, conclusions about the internal structure based on
these wave functions obtained from two-point correlators are not very reliable.

We therefore study an additional observable, the three-point function of the
electromagnetic current between pion states.  This enables us to extract the
electromagnetic vertex of the pion at $T > 0$ and to draw direct conclusions about the
internal structure.  We do this for a medium which has vanishing net baryon
density and a temperature of $T = 0.93\; T_c$.  Our study of the pion vertex function
for space-like photons is the first such investigation of the spatial structure
of a hadron at finite temperature with lattice QCD. For our study we use an
${\cal O}(a)$ improved Wilson action, together with the consistently improved
vector current in order to ensure the absence of $\mathcal{O}(a)$ effects in the
matrix element.

In Sec.~II, we first discuss some general features of the meson form factors at
finite temperature followed by some technical details in Sec.~III. In Sec.~IV,
we present the calculation of the pion two-point function and discuss our
results for screening mass and dispersion relation. The pion
electromagnetic vertex is then extracted from the three-point function and
discussed in detail in Sec.~V. Sec.~VI contains our summary.

\section{Formalism at $T > 0$}
Temperature is introduced in the path integral formalism by restricting the
Euclidean time direction and imposing (anti-) periodic boundary conditions. In
the lattice approach, the temperature is then defined through $T = (N_\tau \;
a)^{-1}$.  
To facilitate the determination of the pion form factor on the lattice, one has
to calculate two observables. These are the two- and three-point Green's
function of an interacting quark-antiquark pair, which carries the quantum
numbers of a pion, at large separation. 
Since at higher temperatures the temporal direction is rather short, it becomes difficult to
reliably filter out the ground state from correlators in the $\tau$-direction
and we use instead spatial correlators in the $z$-direction.

\subsection{The two-point function}

The two point function is given by
\be
G(\bv{x}, \tau) = \langle \phi ( \bv{x}, \tau) \phi^{\dagger} (\bv{0}, 0)\rangle \; ,
\ee
where for definiteness we assume that we are dealing with a $\pi^+$
meson,
\be
\phi^{\dagger}(x) = {\bar \psi}_u(x)\; \gamma^5\; \psi_d(x) \; .
\ee
In the following, we will suppress all flavor, $SU(3)$ color and spin indices.
We want to study the spatial correlator in the $z$-direction,
\be
\label{eq:twopoint}
\tilde{G}(z, \tilde{\bv{p}}) = \int_0^{\frac{1}{T}} d\tau\,\int dx\,\int dy\,
e^{-i \tilde{\bv{p}} \cdot \tilde{\bv{x}}} \;G (\bv{x}, \tau) \; ,
\ee
where 
\be 
\tilde{\bv{x}} = (x,y,\tau) \hspace{1cm} \tilde{\bv{p}} = (p_x, p_y, p_4)
\ee
denote the three-vector parts of the coordinates and momenta
in the so called 'funny space' \cite{Koch:1992nx}.
Due to the periodic boundary conditions for the pion, $p_4$ is
restricted to the Matsubara frequencies,
\be
p_4 = 2\, \pi\, n\, T \equiv \omega_n \; .
\ee
In the following we consider only the lowest contribution, $\omega_0 =
0$, since the next mode at $\omega_1 \approx 1.6$ GeV is already
quite heavy.

\subsubsection{Dispersion relation and wave function renormalisation}

The inverse pion propagator in Euclidean momentum space can be written as
\be
\label{eq:piprop}
\Delta^{-1}(\bv{p}, p_4; T) =  \bv{p}^2 + p_4^{\,2} + m^2 +
\Pi (\bv{p}^2, p_4; T) \; ,
\ee
where $m$ is the bare pion mass and effects due to the presence of the
heat bath are incorporated into the self energy $\Pi$.  In contrast to
the situation at $T=0$, the propagator can now depend separately on
\be
\bv{p}^{\,2} = \bv{p}_{\perp}^{\,2} + p_z^{\,2} \;\;\textrm{and}\;\;p_4 = n \cdot p \, .
\ee
Here, the four-velocity of the heat bath, $n_{\mu}$, is given by
\be
n_\mu = (0,0,0,1) \;.
\ee
In terms of the momentum space propagator in
Eq.~(\ref{eq:piprop}), the spatial correlator in the $z$-direction is given as
\be
\tilde{G} (z, \tilde{\bv{p}}) = \int \; \frac{d p_z}{2\;\pi} e^{-\;i\;p_z\;z}
\; \Delta(\bv{p}, p_4; T) \; .
\ee
Its behavior at large $z$ is determined by the poles with the lowest
$p_z^2$ value of the propagator, Eq. (\ref{eq:piprop}).  For $p_4 =
0$, the poles occur when the spatial momentum satisfies
\be
\label{eq:psq}
\bv{p}^2 = -m^2 - \Pi ( - m_{sc,T}^2, 0; T) = - m^2_{sc,T} \; ,
\ee
where $m_{sc,T}$ denotes the temperature dependent screening mass.
For a given value of the transverse momentum $\bv{p}_{\perp}$, the
pole in $p_z$ is therefore located at $p_{z,0}$, with
\be
p_{z,0}^2 = -  \; E_{sc}^2 (\bv{p}_{\perp}^{\,2}, 0; T)\; ,
\ee
where we have introduced the screening energy,
\be
\label{eq:dispersion}
E_{sc}(\bv{p}_{\perp}^2; T) = \sqrt{\bv{p}_\perp^{\,2} +m_{sc,T}^2} \; .
\ee
The state we filter out for large separation $z$ at a given transverse
momentum $\bv{p}_\perp$ and $p_4 = 0$ is thus the state with the
lowest {\it screening} energy, which satisfies the dispersion relation
Eq.~(\ref{eq:psq}) or Eq.~(\ref{eq:dispersion}).  This state is in
principle different from the ground state of the pion with the lowest
energy. For simplicity, however, we will refer to the state
with the lowest screening energy as the 'ground state'.

In order to obtain the wave function renormalisation for this
state, we expand the propagator around the pole $\bv{p}^2 = -
m_{sc,T}^2$.  Using
\bea
\Pi (\bv{p}^2, 0; T) &\cong& \Pi ( -m_{sc,T}^2 , 0;T) \nonumber \\
&+& (\bv{p}^2 + m_{sc, T}^2)
\; \frac{\partial}{\partial \bv{p}^{\,2}} \Pi(\bv{p}^{\,2}, 0;T) |_{\bv{p}^{\,2} = -m_{sc,T}^2} + \ldots \; ,
\eea
and Eq.~(\ref{eq:psq}), one can write the inverse propagator as
\be
\Delta^{-1} \cong (\bv{p}^2 + m_{sc, T}^2) \left( 1 + \frac{\partial}{\partial \bv{p}^{\,2}}
\Pi(\bv{p}^{\,2}, 0;T) |_{\bv{p}^{\,2} = -m_{sc, T}^2}  + \pi_R (\bv{p}^2) \right) \; ,
\ee
where the remainder vanishes at the pole,
\be
\pi_R (-m_{sc,T}^2) = 0 \; .
\ee
Near the pole, we therefore obtain
\be
\Delta (\bv{p}, p_4 = 0; T) = \frac{Z}{\bv{p}_{\perp}^{\,2} + p_z^{\,2} + m_{sc, T}^2 }\; ,
\ee
where the renormalisation constant $Z$ is defined by
\be
\label{eq:wfrenorm}
Z^{-1} =1 + \frac{\partial}{\partial \bv{p}^{\,2}}
\Pi(\bv{p}^{\,2}, 0;T) |_{\bv{p}^{\,2} = -m_{sc, T}^2} \; .
\ee

\subsection{The three-point function}

The second observable we use is the three-point function. For the pion, using
degenerate quark masses, only connected diagrams contribute. The quark-antiquark
pair propagates from the source at $z_{i} = 0$ to the sink at $z_f$. The photon
couples to the propagating quarks at an intermediate point $z$. In the
continuum, this function reads in the notation introduced in
Eq.~(\ref{eq:twopoint})
\bea
\label{eq:3pt}
G_\mu (z_f, z, \tilde{\bv{p}}_f,  \tilde{\bv{p}}_i) &=& \int^{1/T}\;d^3\tilde{\bv{x}}_f \;
\int^{1/T} \;d^3\tilde{\bv{x}}_i \;e^{- i \tilde{\bv{p}}_f \cdot (\tilde{\bv{x}}_f -
\tilde{\bv{x}}) - i\;\tilde{\bv{p}}_i \cdot \tilde{\bv{x}}} \nonumber \\
&\times&  \langle \phi_R (\tilde{\bv{x}}_f, z_f ) \, j_\mu (\tilde{\bv{x}}, z)\,
\phi^\dagger (\tilde{\bv{0}}, 0)\,\rangle \; ,
\eea
where $j_\mu$ is the quark vector current to which the photon
couples. Details of the current will be given in Section 3.

\subsubsection{Current structure and form factors}
\label{sec:curr_struc}
Given the four-vectors $n, p_i$ and $p_f$, a matrix element of the
electromagnetic current operator has for the pion the general Lorentz
structure,
\bea
J_\mu &=& \langle \pi(\bv{p}_f)|j_{\mu}|\pi(\bv{p}_i\rangle)\nonumber \\ 
&=& e_\pi \{(p_i + p_f)_\mu \;F + q_\mu \;G + n_\mu \;H \}\; ,
\eea
where
\be
q_\mu = (p_f - p_i)_\mu
\ee
is the photon four momentum and $e_\pi$ the pion charge. The functions
$F,G,$ and $H$ are form factors which are functions of the independent
scalar variables and are subject to conditions arising from current
conservation or the Ward identity. In contrast to the situation at
$T=0$ they can depend on more scalar variables, namely
\be
p_i^2, p_f^2, n^2, q^2, n \cdot p_i,\;{\rm and}\;\; n \cdot p_f \; .
\ee
The last two scalars amount to $p_{i,4}$ and $p_{f,4}$.  When
calculating the three-point function, Eq.~(\ref{eq:3pt}), we choose to project
out states with transverse momenta of same magnitude,
\be
\label{eq:perp}
\bv{p}^2_{\perp, f} = \bv{p}^2_{\perp,i} = \bv{p}^2_{\perp} \; .
\ee
By choosing large spatial separations of the vertex from both source
and sink and $p_4 = 0$, we filter out the initial and final pion in
the 'ground state' which satisfies the dispersion relation
Eq.~(\ref{eq:dispersion}).  It is therefore easily seen that the form
factors with this choice of kinematics will only depend on two
scalars, the screening mass $m_{sc,T}$ and
\be
Q^2 = - q^2 = (\bv{p}_{\perp,f} - \bv{p}_{\perp,i})^2 \; .
\ee
In our applications the momentum transfer to the pion, $Q^2$, is
varied by changing the angle between $\bv{p}_{\perp, i}$ and ${\bf
p}_{\perp, f}$ and by choosing different values for ${\bf
p}^2_{\perp}$, Eq.~(\ref{eq:perp}).

Current conservation for matrix elements of the current operator
between initial and final pion states yields in general
\be
0 = (p_f^2 - p_i^2) \;F(m^2_{sc,T}, Q^2) + q^2\; G(m^2_{sc,T}, Q^2)  + n \cdot (p_f - p_i) \;H(m^2_{sc,T}, Q^2) \; .
\ee
Since for our symmetrical kinematics $p^2_f = p^2_i$ and $n \cdot (p_f
- p_i) = 0$, we see that $G$ must vanish. Furthermore, the term
involving $H$ will not contribute if we consider spatial components of
the current. Choosing in particular the $z$-component, we obtain for
the lattice version of the pion current
\bea
\langle\pi(\bv{p}_{\perp,f})\;| j_z|\;\pi(\bv{p}_{\perp,i}) \rangle_{lattice} &=& e_\pi\, F(Q^2, m_{sc, T}^2) \;
\frac{p_{z,f} + p_{z,i}}
{2\;\sqrt{p_{z,f}(\bv{p}^2_{\perp})\;p_{z,i}(\bv{p}^2_{\perp})}} \nonumber \\
&=&  e_\pi\, F(Q^2, m_{sc, T}^2) \; .
\eea
As already pointed out in Ref.~\cite{vanderHeide:2003kh}, the cancellation of the
kinematical factor in the last step due to our symmetric choice of
momenta makes the extraction of the form factor $F$ from the lattice
data more reliable.

\subsubsection{Effective charge}

In comparison to the free pion current at $T=0$, there are two modifications at
$T > 0$. First, the overall renormalisation constant $Z$, Eq.~(\ref{eq:wfrenorm})
at finite temperature, might differ from the corresponding value at $T=0$.
Secondly, the vertex operator gets modified. At the photon point, where $Q^2 =
0$ and $\bv{p}_{\perp,f} = \bv{p}_{\perp,i}$, we absorb both effects into an
effective charge $e_{eff}$,
\be
\langle\pi(\bv{p}_{\perp}) | j_z| \pi(\bv{p}_\perp)\rangle\; \equiv e_{eff}\; 2 \; p_z \; .
\ee

The modification of the vertex can in this case be obtained from the
Ward-Takahashi identity for the vertex operator $j_\mu$,
\be
(p' - p)_\mu \;j_\mu = e_\pi  \{ \Delta^{-1}(\bv{p}', p_{4}'; T) -  \Delta^{-1}(\bv{p}, p_{4};T) \}
\ee
Taking first $\tilde{\bv{p}}' = \tilde{\bv{p}}$ and then the limit
$p_{z}' \rightarrow p_{z}$, one obtains at the photon point
\bea
j_z (Q^2 = 0) &=& e_\pi \;\frac{\partial}{\partial p_z} \Delta^{-1}(\bv{p}, p_4, T) \nonumber \\
 &=& 2\, e_\pi\, p_z\; \left( 1  + \frac{\partial}{\partial \bv{p}^{\,2}}
\Pi(\bv{p}^{\,2}, 0;T)\right)
\eea
When evaluating the matrix element of the vertex operator $j_z$
between states satisfying the dispersion relation
Eq.~(\ref{eq:dispersion}), we have
\bea
j_z (Q^2 = 0)&=& 2\, e_\pi\, p_z\; \left . \left( 1  + \frac{\partial}{\partial \bv{p}^{\,2}}
\Pi(\bv{p}^{\,2}, 0;T) \right)\right |_{\bv{p}^2 = - m^2_{sc,T}} \nonumber \\
&=& 2\, e_\pi\, p_z \; Z^{-1} \; ,
\eea
where $Z$ was defined in Eq.~(\ref{eq:wfrenorm}).  The vertex
correction will thus precisely cancel the wave function
renormalisation factor $Z$, resulting in
\be
e_{eff} = e_\pi \; .
\ee
Thus if we use pion states obtained from spatial correlators, the
effective charge determined from spatial components 
of the conserved current is
the same as the free charge. This complements the remark in
Ref. \cite{Song:1996dg}, where it was noticed that the effective charge
obtained from the $z$-component is different from the one obtained
from the $t$-component if the {\it same} wave function renormalisation
constant is used in both cases. This difference is due to the fact
that Lorentz invariance is broken and the self energy depends
separately on $\bv{p}$ and $p_4$ at $T > 0$.

\section{Technical Details}
Most practical aspects of our lattice calculations are identical to
the procedures in Ref.\cite{vanderHeide:2003kh}. We therefore only give a short
summary.

Our calculations were done on a $N_\sigma \times N_\tau = 32^3 \times
8$ lattice.  The $N_\tau$ value corresponds to $T = 0.93\;T_c$ at the
chosen value of $\beta = 6.0$.  In comparison to Ref.~\cite{vanderHeide:2003kh},
the spatial size of the lattice has been increased from $N_\sigma =
24$ at $T=0$ to $N_\sigma = 32$ to have the same length in the
correlation direction. We generated ${\cal O}(200)$ configurations,
twice as many as at $T=0$, since we expect an increase in fluctuations
in the vicinity of the phase transition.

To facilitate comparison with the results in
\cite{vanderHeide:2003kh}, we used the same five $\kappa$ values,
\be
\label{eq:kaval}
\kappa = 0.13230,\; 0.13330,\; 0.13380,\; 0.13430,\; 0.13480
\ee
which correspond to pion masses of $360\;-\;970\;{\rm MeV}$\footnote{We use $a =
0.105$ fm from \cite{Edwards:1998xf} to set the scale.}  at $T = 0$.  The action
we use is the Sheikholeslami-Wohlert action \cite{Sheikholeslami:1985ij} with
the non-perturbatively determined~\cite{Luscher:1997ug} value $c_{SW} = 1.76923$, which is exact to
order ${\cal O}(a)$.

In order to obtain matrix elements of the current that are correct to
order ${\cal O}(a)$, we have to use the appropriate vector current for
the chosen action. This improved vector current is
\cite{Martinelli:1991ny,Luscher:1997jn, Guagnelli:1998db}
\be
\label{eq:impr}
j_\mu^I = Z_V \{j_\mu^L + a \;c_V\;\partial_\nu\;T_{\mu \nu}\}\; ,
\ee
where
\be
\label{eq:local}
j^L_\mu = {\bar \psi}(x) \;\gamma_\mu\;\psi(x)
\ee
is the local quark current and
\be
T_{\mu \nu} = {\bar \psi}(x)\;\sigma_{\mu \nu}\;\psi(x)\; .
\ee
The renormalisation coefficient $Z_V$ and the improvement constant $c_V$ have
been determined non-perturbatively \cite{Bhattacharya:2000pn}. This improved
current guarantees that on-shell matrix elements only have ${\cal O}(a^2)$
errors. For comparison, we also use the conserved lattice current
\cite{Karsten:1981wd}, in our calculations
\be
\label{eq:cons}
j_\mu^C = {\bar \psi}(x) \;(1 - \gamma_\mu) \; U_\mu (x) \;\psi(x + {\hat \mu})
- {\bar \psi}(x + {\hat \mu}) \;(1 + \gamma_\mu) \; U^\dagger_\mu (x) \;\psi(x) \; ,
\ee
which has ${\cal O}(a)$ discretisation errors away from the forward direction.
To enhance the contribution from the pion ground state at the sink
point $( \bv{\tilde{x}}_f, z_f)$, we use an extended operator $\phi_R$
with a suitably chosen inter-quark distance $R$.  To keep the
calculation gauge invariant, the separated quarks at the
sink are connected by gauge links which are 'fuzzed' to simulate the tube-like nature
of the gluon cloud. In these steps we follow the scheme developed by
Gupta {\it et al.}\cite{Gupta:1993vp} and Lacock {\it et al.} \cite{Lacock:1995qx};
the fuzzed gluon links at the pion sink are created with a link/staple
mixing of $2$ and a fuzzing level of $4$.  As a measure for the
effectiveness of this method, we used the speed with which the
effective screening energy,
\be
\label{eq:effen}
E_{sc,T}^{eff}(z,\tilde{\bv{p}}) = \ln{\frac{\langle\tilde{G}_R(z, \tilde{\bv{p}})\rangle}{\langle \tilde{G}_R(z+1,\tilde{\bv{p}})\rangle}}
\ee
stabilizes for increasing $z$. We varied the quark separation $R$
and found that, just as at $T = 0$, the value $R=3$ was
optimal. The same $R$ was also used for the three-point functions.

\section{Results for the two-point function}
\renewcommand{\arraystretch}{1.2}
\begin{table}
\centering
\caption{Fit results at $T = 0.93 \; T_c$ compared to corresponding data at $T = 0$.}
\subtable[Pion masses.]{
\begin{tabular}{l c c}
\hline
\hline
$\kappa$ & $m_{sc,T}$ & $m_{\pi}$ $(T=0)$\\
\hline
0.13230 & 0.511(3) & 0.516(2)\\
0.13330 & 0.410(4) & 0.414(2)\\
0.13380 & 0.353(4) & 0.356(2)\\
0.13430 & 0.283(5) & 0.287(3)\\
0.13480 & 0.187(8) & 0.194(4)\\
\hline
\hline
\end{tabular}
\label{tab:pion_mass_FT}
}
\subtable[VMD fit parameter $m_V$ compared to the free $\rho$-mass.]{
\begin{tabular}{l c c c}
\hline
\hline
$\kappa$ & $m_V$ & $m_V\; (T=0)$ & $m_{\rho}$ \cite{Gockeler:1998fn} \\
\hline
0.13230 & 0.574(11) & 0.587(19) & 0.623(2)\\
0.13330 & 0.520(13) & 0.528(17) & 0.550(2)\\
0.13380 & 0.495(15) & 0.501(19) & 0.515(3)\\
0.13430 & 0.466(18) & 0.477(21) & 0.485(3)\\
0.13480 & 0.431(25) & 0.454(43) & 0.448(13)\\
\hline
\hline
\end{tabular}
\label{tab:VMD_fit_res}
}
\end{table}
\renewcommand{\arraystretch}{1.0} 

In order to extract information from the lattice data, the two-point function,
Eq.~(\ref{eq:twopoint}), is parameterised, with $p_4 = 0$ as
\be
\label{eq:twostate}
\tilde{G}_R(z, \bv{p}_\bot) = \sum_{n=0}^1 \sqrt{Z_R^n(\bv{p}_{\bot}^2)\; Z_0^n(\bv{p}_{\bot}^2)} \;
e^{- E_{sc}^n(\bv{p}_{\bot}^2)\;N_z /2} \;
\cosh \left[E_{sc}^n(\bv{p}_{\bot}) \left(\frac{N_z}{2} - z)\right) \right] \; ,
\ee
where $N_z = 32$ is the extension of the lattice in the $z$-direction.
The $\cosh$-structure is due to the periodic boundary conditions in
$z$.  The quantities $Z_R^n$ denote the matrix elements $|\langle \Omega |
\phi_R|\;n, {\bv{p}_{\bot}}\rangle |^2\;$, the overlap between the trial
state $\phi_R^{\dagger} |\Omega\rangle$ and the pion state with screening
energy $E_{sc}^n(\bv{p}_{\bot}^2)$ and momentum $\bv{p}_{\bot}$.
Since we use smearing techniques at the sink to suppress higher
excited states, the parameterisation can be restricted to two states.
For a given momentum $\bv{p}_{\bot}$, the parameters we fit for each
state are thus the screening energy, $E_{sc}^n$, and the product of
the amplitudes, $Z^n_R\; Z^n_0$, in Eq.~(\ref{eq:twostate}).
For the momenta, we choose
\be
\bv{p}_{\bot}^2 = 0,\;0.039,\;0.077,\;0.154,\;0.193 \;,
\ee
which are integer multiples of $p_{min}^2$, where $p_{min} =
\frac{2\,\pi}{N_\sigma}$. These values include the momenta used for
our form factor calculation. For each nonzero value of ${\bf
p}_{\bot}^2$ we use several $\bv{p}_{\bot}$ with different directions
in the $\{x,y\}$ plane. On the lattice there is still some rotational
symmetry left and the two-point function is independent of the
direction of the transverse momentum. We therefore average over all
momenta with the same length to increase the stability of the data.

There might still be correlations between different configurations. To deal with 
these, we obtained all our results using the jackknife method
\cite{Quenouille:1949mq, Tukey:1958jt}.

\subsection{Screening masses and dispersion relation}
\begin{figure}
  \psfrag{P2}{$\bv{p}_{\bot}^2$}
  \psfrag{E}{$E_{sc}$}
  \psfrag{m=0.511}{$m_{\pi}=0.511$}
  \psfrag{m=0.410}{$m_{\pi}=0.410$}
  \psfrag{m=0.353}{$m_{\pi}=0.353$}
  \psfrag{m=0.283}{$m_{\pi}=0.283$}
  \psfrag{m=0.187}{$m_{\pi}=0.187$}
  \includegraphics[width=0.75\textwidth]{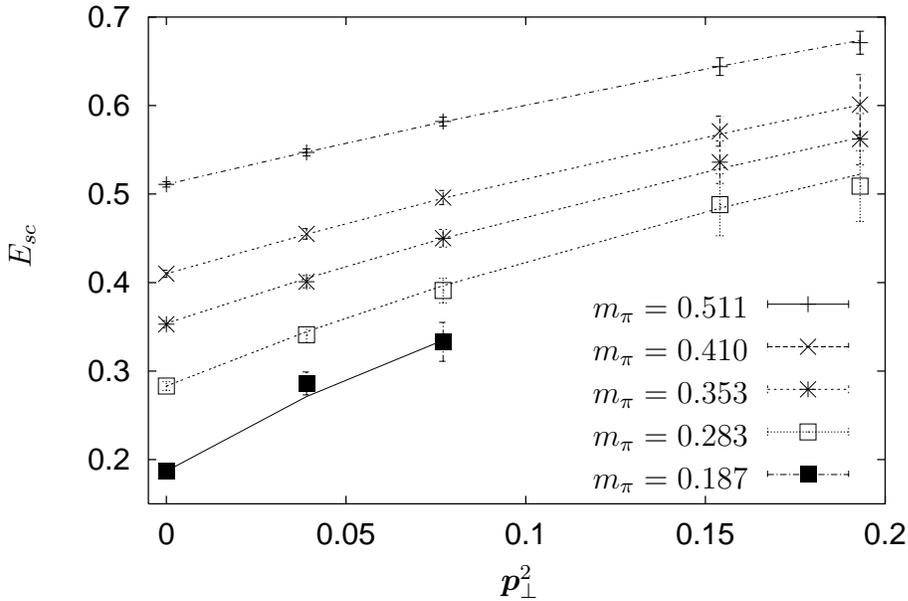}
\caption{$E_{sc}(\bv{p}_{\bot}^2)$ for different pion masses; 
lines: continuum dispersion relation Eq.~(\ref{eq:dispersion}).}
\label{fig:disp_rel}
\end{figure}
The results for the screening masses, $m_{sc,T}$, obtained for $\bv{p}_{\bot} = \bv{0}$,
are given in Table~\ref{tab:pion_mass_FT}.  Also shown in the table are the free
pion masses, $m_\pi$, obtained for the same action at $T=0$ in
Ref. \cite{vanderHeide:2003kh}. As can be seen, the screening masses agree with
the corresponding pole masses at zero temperature within error bars.  This
confirms earlier work \cite{Laermann:2001vg,Laermann:2004kp}, where no significant difference
between screening and zero temperature pole masses were found for a $\pi$ meson
at temperatures below $T_c$; similar observations were also made for the $\rho$
meson.

In order to investigate the dispersion relation for the lattice states, we plot
the screening energies as function of the transverse momentum in
Fig.~\ref{fig:disp_rel}. The screening energies obey the continuum
dispersion relation, Eq.~(\ref{eq:dispersion}), quite well. This nicely confirms
that lattice artifacts are largely suppressed since we are working with an
improved action, which only receives corrections in ${\cal O}(a^2)$. The error
in $E_{sc}(\bv{p}_{\bot})$ grows with increasing momentum and decreasing pion
mass. Due to large fluctuations, reliable results for the lowest
pion mass could not be obtained at the two highest momenta.

\section{The three-point function and the vertex function $F$}
\begin{figure}
\centering
      \psfrag{Extrapolation of T=0}[r][r]{Extrapolation of $T=0$}
      \psfrag{T=0}[r][r]{$T=0$}
      \psfrag{T=0.93 Tc, p2=0.039}[r][r]{$T = 0.93\; T_c$, $\bv{p}_{\bot}^2=0.039$}
      \psfrag{T=0.93 Tc, p2=0.077}[r][r]{$T = 0.93\; T_c$, $\bv{p}_{\bot}^2=0.077$}
      \psfrag{T=0.93 Tc, p2=0.193}[r][r]{$T = 0.93\; T_c$, $\bv{p}_{\bot}^2=0.193$}
      \psfrag{ZV}{$Z_V$}
      \psfrag{mqa}{$m_q\, a$}
      \includegraphics[width=0.75\textwidth]{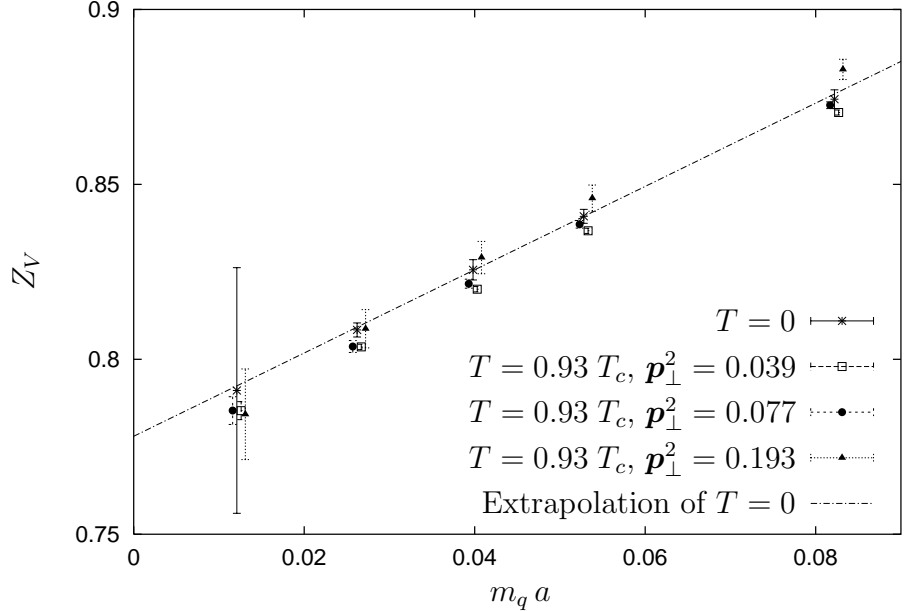}
\caption{Renormalisation constants $Z_V$ as function of quark mass for different $T$ and $p_\bot$.}

\label{fig:renorm_const}
\end{figure}
For the three-point function, we consider a pseudo-scalar source at $z_i = 0$, a
sink at $z_f$, and a coupling of the photon at $0 < z < z_f$.  As was discussed
in Ref.~\cite{vanderHeide:2003kh}, for $T=0$, the most reliable method to
extract the form factor is the use of simultaneous fits of the two- and
three-point functions.  Consequently, this will also be the method used in this
case. In analogy to the zero temperature analysis, we have varied the fit range
in both correlation functions to investigate systematic uncertainties. We found no
significant changes in the ground state parameters.
\begin{figure}
\centering
  \subfigure[$m_{\pi}=0.511$]{
      \psfrag{Q2}[][]{$Q^2$}
      \psfrag{1/F(Q2)}[][]{$1/F(Q^2)$}
      \psfrag{p2 = 0.039}{$\bv{p}_{\bot}^2=0.039$}
      \psfrag{p2 = 0.077}{$\bv{p}_{\bot}^2=0.077$}
      \psfrag{p2 = 0.193}{$\bv{p}_{\bot}^2=0.193$}
      \includegraphics[width=0.45\textwidth]{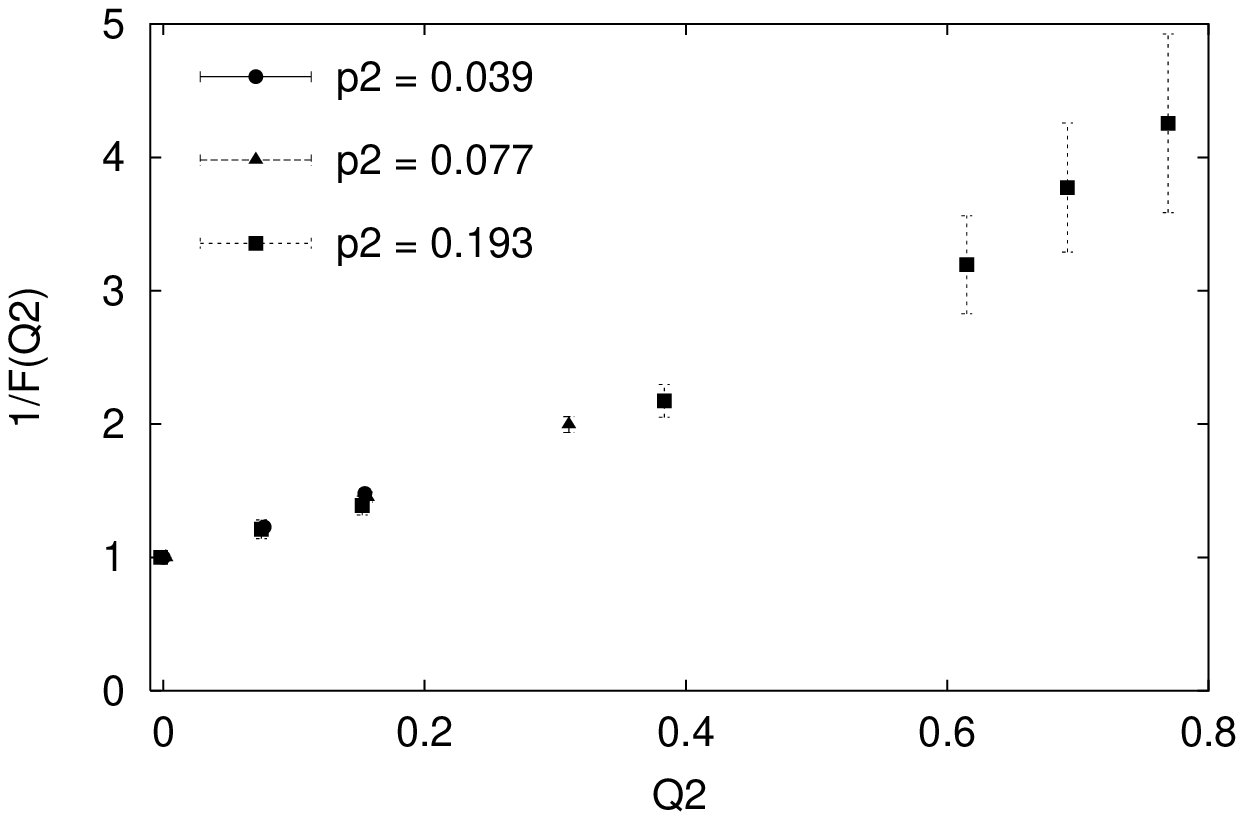}
    \label{fig:FF_VMD_FT_K13230}
  }
  \subfigure[$m_{\pi}=0.283$]{
      \psfrag{Q2}[][]{$Q^2$}
      \psfrag{1/F(Q2)}[][]{$1/F(Q^2)$}
      \psfrag{p2 = 0.039}{$\bv{p}_{\bot}^2=0.039$}
      \psfrag{p2 = 0.077}{$\bv{p}_{\bot}^2=0.077$}
      \psfrag{p2 = 0.193}{$\bv{p}_{\bot}^2=0.193$}
      \includegraphics[width=0.45\textwidth]{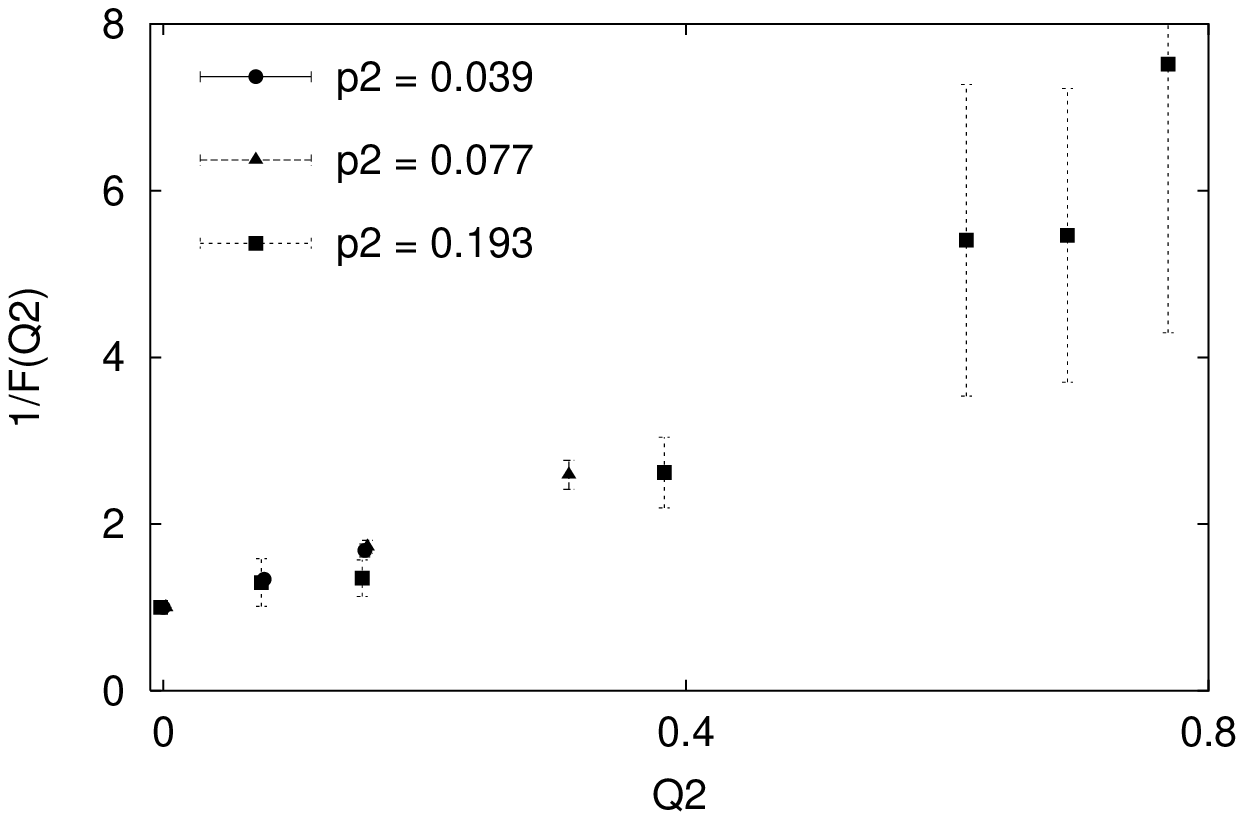}
    \label{fig:FF_VMD_FT_K13430}
  }
  \caption{The inverse of the vertex function $F$ as a function of $Q^2$.} 
\label{fig:FF_VMD_FT}
\end{figure}

We carried out our simulations for the five $\kappa$ values in
Eq.~(\ref{eq:kaval}), corresponding thus to the different screening
masses in Table~\ref{tab:pion_mass_FT}.  Furthermore, we chose three
external momenta, $\bv{p}_{\bot}^2 = 0.039$, $0.077$ and $0.0193$.
As can be seen from our discussion in Sec.~\ref{sec:curr_struc}, we do
not expect a dependence on $\bv{p}_{\bot}^2$ for our specifically chosen kinematical
situation. The data sets for the different transverse momenta will
therefore yield an indication of numerical instabilities in our
results.  For the lowest pion mass and highest $\bv{p}_{\bot}^2$,
fluctuations overwhelmed the data and we did not extract the
parameters of the three-point function.

As shown above, for our kinematics and by considering the
$z$-component of the current, the resulting matrix element takes on a
very simple structure with only one single vertex function $F$, which
depends on $Q^2$ and the screening mass.  In our calculations, we have
considered three choices of the current operator on the lattice: the
conserved current, Eq.~(\ref{eq:cons}), the renormalised local
current, Eq. (\ref{eq:local}), and the improved current,
Eq. (\ref{eq:impr}).  At $Q^2 = 0$, for periodic boundary
conditions, the three-point function calculated
with a conserved current should satisfy the relation \cite{Barad:1984px}
\be
\label{eq:second}
\frac{G_3(z_f, z; \tilde{\bv{p}}, \tilde{\bv{p}}) - G_3(z_f, z', \tilde{\bv{p}}, \tilde{\bv{p}})}
{G(z_f, \tilde{\bv{p}})} = 1 \; ,
\ee
where $z_f < z' < N_\sigma$. In this form, the current operator is
inserted twice to measure two separate contributions to the total
charge reaching the sink at $z_f$. The first takes into account the
charge from the source reaching the sink by passing through $z$.  The
'second insertion' at $z'$ accounts for the charge that leaves the
source in the negative $z$-direction and arrives at the sink by
passing through $z'$, which is possible due to the periodic boundary
conditions.  We have confirmed that for the conserved current
Eq.~(\ref{eq:second}) is satisfied to high accuracy, typically to
${\cal O}(10^{-4})$.

For both the local and the improved current we need the overall renormalisation
constant $Z_V$,
\be
Z_V= Z_V^0 (1\, + \, b_V\, m_q).
\ee
We have determined it by demanding that Eq.~(\ref{eq:second}) is
satisfied. Since the additional tensor term in the improved current is a total
divergence, it does not contribute to the total charge and therefore $Z_V$ is
identical for the local and improved current.  Fig.~\ref{fig:renorm_const} shows
the $Z_V$-values we obtain for several $\kappa$ and $\bv{p}_{\bot}^2$
values. The higher momenta are more problematic for the extraction of the vertex
functions. They have larger error bars and the extracted $Z_V$
values tend to lie slightly higher than the lower momentum values. As the
figure shows, the $T = 0.93\;T_c$ results deviate by less than $2\%$ from $T =
0$ values extracted in Ref.~\cite{vanderHeide:2003kh}. They also have the same
linear dependence on the quark mass.  Thus, our values for $Z_V^0$ as well
as $b_V$ are in very good agreement with the non-perturbative determination of
Bhattacharya {\it et al.}  \cite{Bhattacharya:2000pn}. In the following, we
therefore use their value for the small improvement constant $c_V$ for the
vector current, which contributes to the vertex only away from the forward
direction. 

\subsection{Results for the vertex function $F$}
\begin{figure}
\centering
      \psfrag{Q2}[][]{$Q^2$}
      \psfrag{F(Q2)}[][]{$F(Q^2)$}
      \psfrag{mpi = 0.511}{$m_{\pi}=0.511$}
      \psfrag{mpi = 0.410}{$m_{\pi}=0.410$}
      \psfrag{mpi = 0.353}{$m_{\pi}=0.353$}
      \psfrag{mpi = 0.283}{$m_{\pi}=0.283$}
      \psfrag{mpi = 0.187}{$m_{\pi}=0.187$}
      \includegraphics[width=0.75\textwidth]{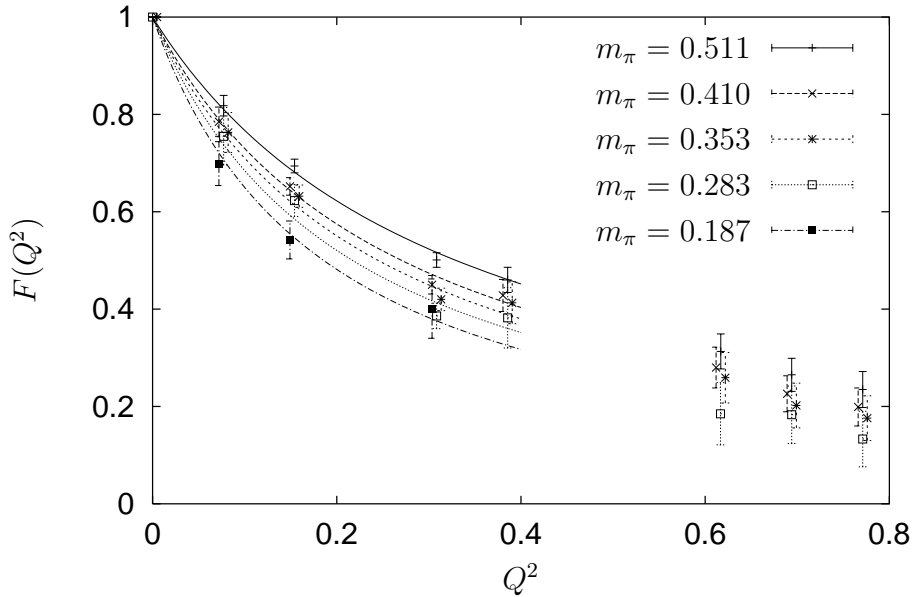}
\caption{The vertex function for different pion masses. Curves: fits to the VMD model.}
\label{fig:FF_diff_kappa_FT}
\end{figure}
We now discuss the values for the vertex function we extracted with the improved
current for different photon momenta $Q^2$.  At $T = 0$, it was found that a
monopole parameterisation of the pion charge form factor described the obtained
results quite well for the lower $Q^2$ values, a feature which can be explained
by the vector meson dominance (VMD) model.  We investigated if the analogous
parameterisation,
\be
\label{eq:monopole}
F(Q^2, m_{sc, T}^2) = [ 1 + \frac{Q^2}{m_V^2}]^{-1} \; ,
\ee
where $m_V$ is a fit parameter, also yields a useful description for the
vertex function we extract at $T > 0$. We therefore plot $1/F$, which
should be a straight line for the parameterisation to work.

As can be seen in Fig.~\ref{fig:FF_VMD_FT} for two values of the pion (screening)
mass, our numerical results show that for $Q^2$ up to $0.4$ a monopole
fit will work quite well.  At higher $Q^2$, the lighter mass shows a
stronger deviation from a straight line, but with larger error bars as
well. When comparing data from the different transverse momenta sets,
we see that the higher $\bv{p}_{\bot}^2$ values have larger error bars and
for the lighter mass are more scattered around a straight line.
\begin{figure}[!t]
\centering
\subfigure[$m_{\pi}= 0.410$]{
      \psfrag{Q2}[][]{$Q^2$}
      \psfrag{F(Q2)}[][]{$F(Q^2)$}
      \psfrag{mpi=0.410}{}
      \psfrag{T = 0.93 Tc}[][]{$T=0.93 \; T_c$}
      \psfrag{T = 0}[][]{$T=0$}
      \includegraphics[width=0.75\textwidth]{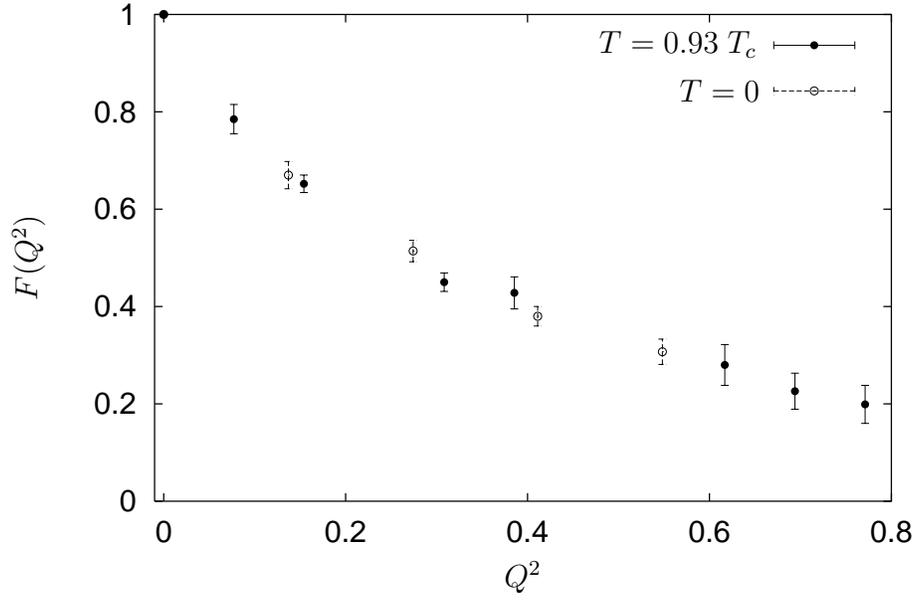}
      \label{fig:FF_K13330_diff_T_comp}
}
\subfigure[$m_{\pi}=0.283$]{
      \psfrag{Q2}[][]{$Q^2$}
      \psfrag{F(Q2)}[][]{$F(Q^2)$}
      \psfrag{mpi=0.283}{}
      \psfrag{T = 0.93 Tc}[][]{$T=0.93 \; T_c$}
      \psfrag{T = 0}[][]{$T=0$}
      \includegraphics[width=0.75\textwidth]{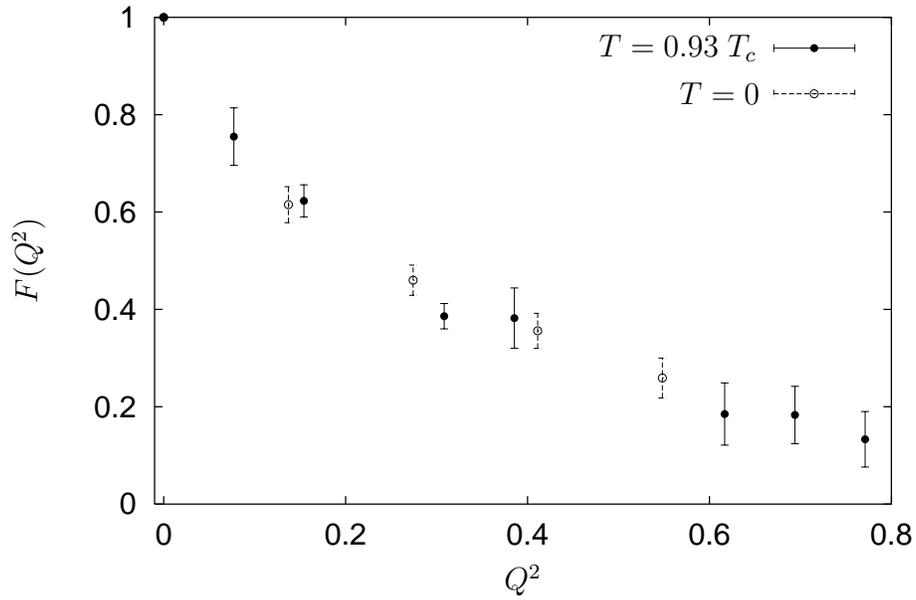}
      \label{fig:FF_K13430_diff_T_comp}
}
  \caption{The vertex function F compared with the form factor at $T=0$.}
\label{fig:FF_diff_T}
\end{figure}

Fig.~\ref{fig:FF_diff_kappa_FT} shows all our results, appropriately averaged
over the different $\bv{p}_{\bot}^2$ values, together with the monopole
fits. For the lightest pion mass, the fluctuations for the higher
$\bv{p}_{\bot}^2$ values were too strong to allow a reliable extraction of the
vertex function at large $Q^2$ values.  As expected, $F(Q^2)$ drops off more
rapidly as the pion mass decreases. The fits deviate more strongly from the data
at higher $Q^2$ for the lighter pions. As shown in Table~\ref{tab:VMD_fit_res},
the extracted parameter $m_V$ agrees quite well with the corresponding fit
parameter extracted at $T = 0$.  Both lie close to the free $\rho$-mass obtained
from lattice QCD in Ref.~\cite{Gockeler:1998fn}. This agreement with $m_\rho$
gets better as the pion mass decreases towards the physical value. Our fit at $T
= 0.93 \;T_c$ thus in general supports the simple VMD picture for the low $Q^2$
data also at a temperature just below $T_c$.  A more detailed look at the
function $F$ is taken in Fig.~\ref{fig:FF_diff_T}. The two different pion masses
show how the difficulties to extract information about $F$ increase as the pion
mass decreases. However, within the error bars, both examples again show that
there is no overall significant difference to the form factor $F(Q^2)$ we
extracted at $T = 0$. This is in contrast to most theoretical expectations based
on effective models. In terms of the mass parameter $m_V$ of our monopole fit,
our results therefore do not support a significant dropping of the vector meson
mass as $T$ increases to $0.93\;T_c$, at least not for the state with the lowest
screening mass we project out by our spatial correlators.

\begin{figure}
\centering
      \psfrag{<r2> (fm2)}[][]{$\langle r^2 \rangle$ (fm$^2$)}
      \psfrag{mpi2 (GeV2)}[][]{$m_{\pi}^2$ (GeV$^2$)}
      \psfrag{0.93 Tc}[r][r]{$T=0.93 \; T_c$}
      \psfrag{T=0}[r][r]{$T=0$}
      \psfrag{VMD}[][]{VMD}
      \psfrag{Exp. value (T=0)}[r][r]{Exp. value $(T=0)$}
      \includegraphics[width=0.75\textwidth]{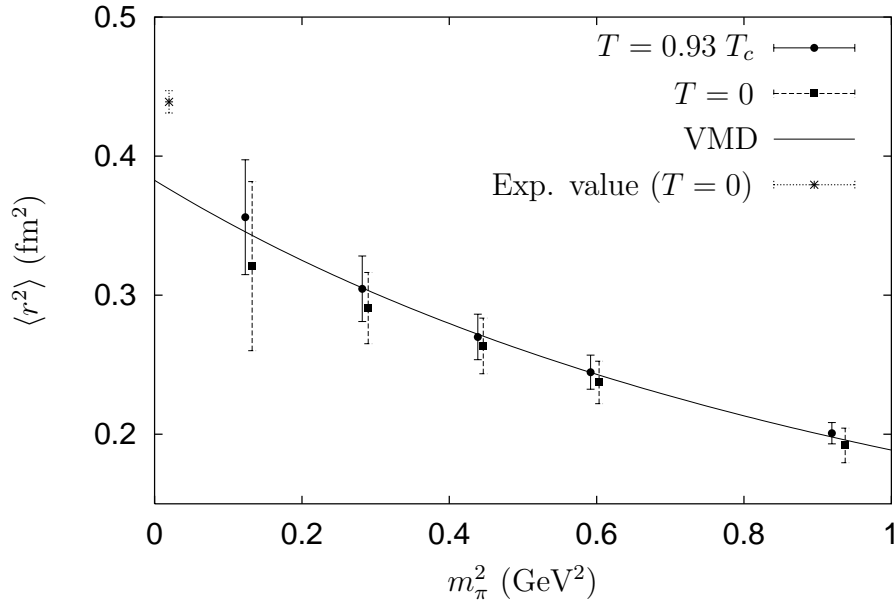}
      \caption{The pion 'radius' at $0.93 \; T_c$ as a function of its mass. }
\label{fig:rms_FT}
\end{figure}

By identifying the vertex function $F(Q^2, m_{sc,T})$ at $T = 0.93 \; T_c$ with
the form factor of a pion embedded in a heat bath, we can also translate our
findings into statements about its spatial extension and mean square
radius. Fig.~\ref{fig:rms_FT} shows $\langle r^2 \rangle$ as obtained from the
slope of $F$ at $Q^2 = 0$, using the VMD parameterisation. Both the $T= 0$ and $T
= 0.93\;T_c$ data lead to the same picture. There is no significant difference
between the value for $\langle r^2\rangle$ and in the dependence on the pion
mass for both temperatures. Using the VMD parametrisation and extrapolating the
'$\rho$' mass $m_V$, linearly in $m_{\pi}^2$, to the physical limit, we arrive at a
value for $\langle r^2 \rangle$ that is about $15\%$ below the experimental
value for a free pion \cite{Amendolia:1986wj}. This small discrepancy could be
due to the quenched approximation. However, Alexandrou {\it et
al.}~\cite{Alexandrou:2002nn} have calculated density-density correlations in
quenched as well as unquenched QCD at $T=0$ and found only rather small effects
for the pion. Recent unquenched calculations of the free pion form factor itself can be
found in \cite{Brommel:2005ee,Hashimoto:2005am}.

\section{Summary}

In this paper, we have 
presented the first investigation of
the electromagnetic vertex of a pion at finite temperature
by means of numerical simulation. The temperature was chosen as
$0.93\; T_c$, just below the critical temperature for the
transition to the deconfined phase of QCD.

We considered two- and three-point functions for the pion for various transverse
momenta.  As the temporal extent of the lattice is limited by the inverse
temperature we have calculated spatial correlators in the $z$-direction to study
the dispersion relation and to extract the vertex functions of a pion embedded
in a heat bath.  For several quark masses we confirmed earlier findings that
within errors the screening mass at $0.93\;T_c$ is the same as the free pion
mass calculated on the lattice.  Furthermore, except for the lightest pion and
the largest momentum where errors are large, we could show that the dispersion
relation for the lattice results was sufficiently well described by the
continuum expression.

For the symmetrical kinematical conditions specifically chosen by us, 
the electromagnetic vertex
of the pion can be described in terms of only a single vertex function
$F(Q^2,\;m_{sc})$, which depends on the square of the photon four-momentum and
the screening mass at the chosen temperature. For $T = 0$, this function is
equal to the free pion form factor $F(Q^2)$. Within error bars, we
found that for all quark masses and all $Q^2$ we
considered, the vertex function agreed with the corresponding form
factor at $T = 0$. Furthermore, we showed that at low $Q^2$ the vertex function
for $T = 0.93\;T_c$ is well described by a monopole fit as
suggested by a vector meson dominance model.  The mass parameter $m_V$ we
extracted equals the fit parameter found earlier at $T = 0$ within errors. 
As the quark mass was decreased, this parameter was seen to
approach the lattice value $m_\rho$ for a $\rho$-meson at $T=0$.
The simple vector meson dominance model thus remains a
good description of the low $Q^2$ results also at a temperature
close to $T_c$.

By interpreting the vertex function as the form factor of the pion state with
the lowest screening mass, we can also translate our low $Q^2$ results into a
mean square radius $<r^2>$. Contrary to most predictions based on effective
models that the radius of a pion will gradually grow as the temperature of the
medium increases, we did not find significant differences from the situation at $T =
0$.

To conclude, our results for the screening masses at $T = 0.93\;T_c$ confirm
what earlier studies of the pion had already indicated. Our main new finding is
that there is also no significant change when one looks at the vertex function
$F$, a quantity that is directly sensitive to the internal structure of the
pion. It will of course be interesting to continue these studies of the
three-point functions to higher temperatures to see whether significant changes
occur when $T_c$ is crossed. Such studies are presently in progress
\cite{Heide:2006kl}.


\begin{thebibliography}{10}

\bibitem{Schulze:1994fy}
H.~J. Schulze,
\newblock J. Phys. {\bf G20}, 531 (1994).

\bibitem{Dominguez:1996kf}
C.~A. Dominguez, M.~S. Fetea, and M.~Loewe,
\newblock Phys. Lett. {\bf B387}, 151 (1996), hep-ph/9608396.

\bibitem{Song:1996dg}
C.~Song and V.~Koch,
\newblock Phys. Rev. {\bf C54}, 3218 (1996), nucl-th/9608010.

\bibitem{GomezNicola:2004gg}
A.~Gomez~Nicola, F.~J. LLanes-Estrada, and J.~R. Pelaez,
\newblock Phys. Lett. {\bf B606}, 351 (2005), hep-ph/0405273.

\bibitem{Karsch:2003jg}
F.~Karsch and E.~Laermann,
\newblock (2003), hep-lat/0305025,
\newblock Published in *Quark gluon plasma III, Hwa, R.C. (ed.) \textit{et
  al.}* 1-59.

\bibitem{vanderHeide:2003ip}
J.~van~der Heide, M.~Lutterot, J.~H. Koch, and E.~Laermann,
\newblock Phys. Lett. {\bf B566}, 131 (2003), hep-lat/0303006.

\bibitem{vanderHeide:2003kh}
J.~van~der Heide, J.~H. Koch, and E.~Laermann,
\newblock Phys. Rev. {\bf D69}, 094511 (2004), hep-lat/0312023.

\bibitem{Koch:1992nx}
V.~Koch, E.~V. Shuryak, G.~E. Brown, and A.~D. Jackson,
\newblock Phys. Rev. {\bf D46}, 3169 (1992), hep-ph/9204236.

\bibitem{Edwards:1998xf}
R.~G. Edwards, U.~M. Heller, and T.~R. Klassen,
\newblock Nucl. Phys. {\bf B517}, 377 (1998), hep-lat/9711003.

\bibitem{Sheikholeslami:1985ij}
B.~Sheikholeslami and R.~Wohlert,
\newblock Nucl. Phys. {\bf B259}, 572 (1985).

\bibitem{Luscher:1997ug}
M.~L{\"u}scher, S.~Sint, R.~Sommer, P.~Weisz, and U.~Wolff,
\newblock Nucl. Phys. {\bf B491}, 323 (1997), hep-lat/9609035.

\bibitem{Martinelli:1991ny}
G.~Martinelli, C.~T. Sachrajda, and A.~Vladikas,
\newblock Nucl. Phys. {\bf B358}, 212 (1991).

\bibitem{Luscher:1997jn}
M.~L{\"u}scher, S.~Sint, R.~Sommer, and H.~Wittig,
\newblock Nucl. Phys. {\bf B491}, 344 (1997), hep-lat/9611015.

\bibitem{Guagnelli:1998db}
M.~Guagnelli and R.~Sommer,
\newblock Nucl. Phys. Proc. Suppl. {\bf 63}, 886 (1998), hep-lat/9709088.

\bibitem{Bhattacharya:2000pn}
T.~Bhattacharya, R.~Gupta, W.-J. Lee, and S.~R. Sharpe,
\newblock Phys. Rev. {\bf D63}, 074505 (2001), hep-lat/0009038.

\bibitem{Karsten:1981wd}
L.~H. Karsten and J.~Smit,
\newblock Nucl. Phys. {\bf B183}, 103 (1981).

\bibitem{Gupta:1993vp}
R.~Gupta, D.~Daniel, and J.~Grandy,
\newblock Phys. Rev. {\bf D48}, 3330 (1993), hep-lat/9304009.

\bibitem{Lacock:1995qx}
UKQCD, P.~Lacock, A.~McKerrell, C.~Michael, I.~M. Stopher, and P.~W.
  Stephenson,
\newblock Phys. Rev. {\bf D51}, 6403 (1995), hep-lat/9412079.

\bibitem{Gockeler:1998fn}
M.~G{\"o}ckeler {\em et~al.},
\newblock Phys. Rev. {\bf D57}, 5562 (1998), hep-lat/9707021.

\bibitem{Quenouille:1949mq}
M.~H. Quenouille,
\newblock J. Roy. Statist. Soc. B {\bf B11}, 18 (1949).

\bibitem{Tukey:1958jt}
J.~W. Tukey,
\newblock Ann. Math. Stat. {\bf 29}, 614 (1958).

\bibitem{Laermann:2001vg}
E.~Laermann and P.~Schmidt,
\newblock Eur. Phys. J. {\bf C20}, 541 (2001), hep-lat/0103037.

\bibitem{Laermann:2004kp}
E.~Laermann {\em et~al.},
\newblock Proceedings of SEWM04 , 206.

\bibitem{Barad:1984px}
K.~Barad, M.~Ogilvie, and C.~Rebbi,
\newblock Phys. Lett. {\bf B143}, 222 (1984).

\bibitem{Amendolia:1986wj}
NA7, S.~R. Amendolia {\em et~al.},
\newblock Nucl. Phys. {\bf B277}, 168 (1986).

\bibitem{Alexandrou:2002nn}
C.~Alexandrou, P.~de~Forcrand, and A.~Tsapalis,
\newblock Phys. Rev. {\bf D66}, 094503 (2002), hep-lat/0206026.

\bibitem{Brommel:2005ee}
D.~Brommel {\em et~al.},
\newblock PoS {\bf LAT2005}, 360 (2005), hep-lat/0509133.

\bibitem{Hashimoto:2005am}
JLQCD, S.~Hashimoto {\em et~al.},
\newblock PoS {\bf LAT2005}, 336 (2005), hep-lat/0510085.

\bibitem{Heide:2006kl}
J.~van~der Heide, J.~H. Koch, and E.~Laermann,
\newblock In preparation.

\end{thebibliography}

\end{document}